\documentclass[epj]{webofc}
\usepackage[varg]{txfonts}   
\newcommand{\bea}{\begin{eqnarray}}
\newcommand{\eea}{\end{eqnarray}}
\newcommand{\be}{\begin{equation}}
\newcommand{\ee}{\end{equation}}
\newcommand{\tr}{\mathop{\mathrm{Tr}}}

\newcommand{\p}{{\bf p}}

\newcommand{\slparc}{\mbox{$\,\slash$ \hspace{-0.7em}$\partial$}}

\newcommand{\nom}{{\nonumber}}

\newcommand{\expm}{e^{-\beta\left(E_p-\mu_q\right)}}

\newcommand{\expmmm}{e^{-3\beta\left(E_p-\mu_q\right)}}
\newcommand{\expp}{e^{-\beta\left(E_p+\mu_q\right)}}
\newcommand{\expppp}{e^{-3\beta\left(E_p+\mu_q\right)}}

\allowdisplaybreaks

\woctitle{MESON2014 - the 13$^\textrm{th}$ International Workshop on Meson Production, Properties and Interaction}
\begin{document}
\selectlanguage{english}
\title{Effects of (axial)vector mesons on the chiral phase transition:
initial results}

\author{P. Kov{\'a}cs\inst{1}\fnsep\thanks{\email{kovacs.peter@wigner.mta.hu}} \and
        Zs. Sz{\'e}p\inst{2} \and
        Gy. Wolf\inst{1} 
}

\institute{Institute for Particle and Nuclear Physics, Wigner Research
  Center for Physics, Hungarian Academy of Sciences, H-1525 Budapest,
  Hungary 
  \and
  MTA-ELTE  Statistical and Biological Physics Research Group, H-1117 Budapest,
  Hungary
}

\abstract{%
  We investigate the effects of (axial)vector mesons on the chiral phase
  transition in the framework of an SU(3), (axial)vector meson extended
  linear sigma model with additional constituent quarks and Polyakov
  loops. We determine the parameters of the Lagrangian at zero
  temperature in a hybrid approach, where we treat the mesons at
  tree-level, while the constituent quarks at 1-loop level. We assume
  two nonzero scalar condensates and together with the Polyakov-loop
  variables we determine their temperature dependence according to the
  1-loop level field equations.

}
\maketitle
\section{Introduction}
\label{intro}
\vspace{-0.1cm}
Nowadays, investigation of the QCD phase diagram is a very important
subject both theoretically and experimentally. Ongoing and upcoming
heavy ion experiments such as RHIC, CERN LHC and CBM FAIR explore
different regions of the QCD phase space. Since properties of the
phase space/boundary is still not settled
theoretically/experimentally, it is worth to investigate this subject
thoroughly.

Our starting point is the (axial)vector meson extended linear sigma
model with additional constituent quarks and Polyakov-loop
variables. The previous version of the model, without constituent
quarks and Polyakov-loops, was exhaustively analyzed at zero
temperature in \cite{Parganlija_2013}\footnote{In the present work we
  use a different anomaly term ($c_1$ term). This, however, does not
  influence the results much.}. The Lagrangian of the model is given
by,
\begin{align}
  \mathcal{L} & = \tr[(D_{\mu}\Phi)^{\dagger}(D_{\mu}\Phi)]-m_{0}%
  ^{2}\tr(\Phi^{\dagger}\Phi)-\lambda_{1}[\tr(\Phi^{\dagger}%
  \Phi)]^{2}-\lambda_{2}\tr(\Phi^{\dagger}\Phi)^{2}\nom \\
  & -\frac{1}{4}\tr(L_{\mu\nu}^{2}+R_{\mu\nu}^{2})+\tr\left[ \left(
      \frac{m_{1}^{2}}{2}+\Delta\right)
    (L_{\mu}^{2}+R_{\mu}^{2})\right]
  +\tr[H(\Phi+\Phi^{\dagger})]\nom \\
  & +c_{1}(\det\Phi+\det\Phi^{\dagger})+i\frac{g_{2}}{2}(\tr
  \{L_{\mu\nu}[L^{\mu},L^{\nu}]\}+\tr\{R_{\mu\nu}[R^{\mu},R^{\nu
  }]\})\nom \\
  & +\frac{h_{1}}{2}\tr(\Phi^{\dagger}\Phi)\tr(L_{\mu}
  ^{2}+R_{\mu}^{2})+h_{2}\tr[(L_{\mu}\Phi)^{2}+(\Phi R_{\mu}
  )^{2}]+2h_{3}\tr(L_{\mu}\Phi R^{\mu}\Phi^{\dagger})\\
  & +g_{3}[\tr(L_{\mu}L_{\nu}L^{\mu}L^{\nu})+\tr(R_{\mu}R_{\nu
  }R^{\mu}R^{\nu})] + g_{4}[\tr\left( L_{\mu}L^{\mu}L_{\nu}L^{\nu
    }\right) \nom\\
    & + \tr\left( R_{\mu}R^{\mu}R_{\nu}R^{\nu}\right)] + g_{5}\tr\left( L_{\mu}L^{\mu}\right) \,\tr\left(
    R_{\nu}R^{\nu}\right) + g_{6} [\tr(L_{\mu}L^{\mu})\,\tr(L_{\nu}L^{\nu})\nom \\
    & + \tr(R_{\mu}R^{\mu})\,\tr(R_{\nu}R^{\nu})] + \bar{\Psi}i
      \slparc\Psi - g_{F}\bar{\Psi}\left(\Phi_{S} +
        i\gamma_5\Phi_{PS}\right)\Psi,\nom 
\end{align}
where $D^{\mu}\Phi = \partial^{\mu}\Phi-ig_{1}(L^{\mu}\Phi-\Phi
R^{\mu})-ieA_{e}^{\mu}[T_{3},\Phi],$ $L^{\mu\nu}
= \partial^{\mu}L^{\nu}-ieA_{e}^{\mu}[T_{3},L^{\nu}] -
\left\{\partial^{\nu}L^{\mu} - ieA_{e}^{\nu}[T_{3},L^{\mu}]\right\},$
and $R^{\mu\nu} = \partial^{\mu}R^{\nu} - ieA_{e}^{\mu}[T_{3},R^{\nu}]
- \left\{ \partial^{\nu}R^{\mu}-ieA_{e}^{\nu}[T_{3},R^{\mu}]\right\}.$
Here $\Phi$ stands for the scalar and pseudoscalar fields, $L^{\mu}$
and $R^{\mu}$ for the left and right handed vector fields, $\Psi=(u,
d, s)^{\text{T}}$ for the constituent quark fields, while $H$ for the
external field.  

\section{Parametrization}
\label{sec-param}

In order to go to finite temperature/chemical potential, parameters of
the Lagrangian have to be determined, which is done at $T=\mu=0$. For
this we calculate tree-level masses and decay widths of the model and
compare them with the experimental data taken from the PDG
\cite{PDG}. For the comparison we use a $\chi^2$ minimalization method
\cite{MINUIT} to fit our parameters (for more details see
\cite{Parganlija_2013}). It is important to note that in the present
work we also included in the scalar and pseudoscalar masses the
contributions coming from the fermion vacuum fluctuations by adapting
the method of \cite{Chatterjee:2011jd}. 

We have $14$ unknown parameters, namely $m_0$, $\lambda_1$,
$\lambda_2$, $c_1$, $m_1$, $g_1$, $g_2$, $h_1$, $h_2$, $h_3$,
$\delta_S$, $\Phi_N$, $\Phi_S$, and $g_F$. Here $g_F$ is the coupling
of the additionally introduced Yukawa term, which can be determined
from the constituent quark masses through the equations $m_{u/d} =
g_F\phi_N/2$, $m_s = g_F\phi_s/\sqrt{2}$. 

It is worth to note that if we do not consider the very uncertain
scalar-isoscalar sector $m_0$, and $\lambda_1$ always appear in the
same combination $C_{1} = m_{0}^{2} + \lambda_{1} \left(\phi_{N}^{2}
  +\phi_S^2\right)$ in all the expressions, thus we can not determine
them separately. Additionally a similar combination appears for $m_1$
and $h_1$ in the vector sector as $C_{2} = m_{1}^{2} + \frac{h_{1}}{2}
\left(\phi_{N}^{2} + \phi_{S}^{2}\right)$ (see details in
\cite{Parganlija_2013}). The parameter values are given in
Table~\ref{tab-param}.
\begin{table}
\centering
\caption{Parameters determined by $\chi^2$ minimalization}
\label{tab-param}   
\begin{tabular}{llll}
\hline
Parameter & Value & Parameter & Value \\\hline
$\phi_{N}$ [GeV]& $0.1622 $ & $h_{2}$ & $11.6586 $ \\
$\phi_{S}$ [GeV]& $0.1262 $ & $h_{3}$ & $4.7028 $ \\
$C_{1}$ [GeV$^2$] & $-0.7537 $ & $\delta_{S}$ [GeV$^2$] & $0.1534 $ \\
$C_{2}$ [GeV$^2$] & $0.3953 $ & $c_{1}$ [GeV]& $1.12 $ \\
$\lambda_{1}$ & undetermined &  $g_{1}$ & $-5.8943$ \\
$\lambda_{2}$ & $65.3221 $ & $g_{2}$ & $-2.9960$ \\
$h_{1}$ & undetermined & $g_{F}$ & $4.9429$ \\\hline
\end{tabular}
\end{table}
Since $\lambda_1$ is undetermined it can be tuned to change the
$f_0^L$ (a.k.a. $\sigma$) mass, which has, as we will see, a huge
effect on the thermal properties of the model.

\section{Field equations}
\label{sec-eqn}

In our approach we have four order parameters, which are the $\phi_N$
non-strange and $\phi_S$ strange condensates, and the $\Phi$ and
$\bar{\Phi}$ Polyakov-loop variables. The condensates arise due to the
spontaneous symmetry breaking\footnote{Since isospin symmetry is
  assumed, we have only two condensates: $\phi_N$ and $\phi_S.$}, while
the Polyakov-loop variables naturally emerge in mean field approximation,
if one calculates free fermion grand canonical potential on a constant
gluon background. The effect of fermions propagating on a constant
gluon background in the temporal direction formally amounts to the
appearance of imaginary color dependent chemical potentials (for
details see \cite{Marko_2010, Korthals_1999}).

At finite temperature/baryochemical potential we can set up four
coupled field equations for the four fields, which are just the
requirements that the first derivatives of the grand canonical
potential according to the fields must vanish. As a first
approximation we apply a hybrid approach in which we only consider
vacuum and thermal fluctuations for the fermions, but not for the
bosons. 
Within this simplified treatment the equations are the following
\begin{align}
-\frac{d }{d \Phi}\left( \frac{U(\Phi,\bar\Phi)}{T^4}\right)
+ \frac{2 N_c}{T^3}\sum_{q=u,d,s} \int \frac{d^3 \p}{(2\pi)^3}
 \left(\frac{e^{-\beta E_q^{-}(p)}}{g_q^-(p)} +  \frac{e^{-2\beta E_q^{+}(p)}}{g_q^+(p)}
\right) &= 0,\label{eq_Phi}\\
-\frac{d}{d \bar\Phi}\left( \frac{U(\Phi,\bar\Phi)}{T^4}\right)
+ \frac{2 N_c}{T^3}\sum_{q=u,d,s}  \int \frac{d^3 \p}{(2\pi)^3}
 \left(\frac{e^{-\beta E_q^{+}(p)}}{g_q^+(p)} +  \frac{e^{-2\beta E_q^{-}(p)}}{g_q^-(p)}
\right) &= 0,\label{eq_Phibar}\\
m_0^2 \phi_N + \left(\lambda_1 + \frac{1}{2} \lambda_2 \right)
\phi_N^3 + \lambda_1 \phi_N \phi_S^2 - h_N
+\frac{g_F}{2}N_c\left(\langle u{\bar u}\rangle_{_{T}} + \langle d{\bar
  d}\rangle_{_{T}} \right) &= 0,\label{eq_phiN}\\
m_0^2 \phi_S + \left(\lambda_1 + \lambda_2 \right)
\phi_S^3 + \lambda_1 \phi_N^2 \phi_S - h_S
+\frac{g_F}{\sqrt{2}}N_c \langle s{\bar s}\rangle_{_{T}} &= 0,\label{eq_phiS}
\end{align}
where
\begin{align*}
g_q^+(p) &= 1 + 3\left( \bar\Phi + \Phi e^{-\beta E_q^{+}(p)} \right)
e^{-\beta E_q^{+}(p)} + e^{-3\beta E_q^{+}(p)}, \\
g_q^-(p) &= 1 + 3\left( \Phi + \bar\Phi e^{-\beta E_q^{-}(p)} \right)
e^{-\beta E_q^{-}(p)} + e^{-3\beta E_q^{-}(p)}, \\
E_q^{\pm}(p) =& E_q(p) \mp \mu_B/3,\; E_{u/d}(p) = \sqrt{p^2 +
  m_{u/d}^2},\; E_{s}(p) = \sqrt{p^2 + m_{s}^2}, \\
\end{align*}
and
\begin{align}
\langle q{\bar q}\rangle_{_{T}} &= -4m_q \int \frac{d^3
  \p}{(2\pi)^3}\frac{1}{2E_q(p)}\left(1 - f^-_\Phi(E_q(p)) -
  f^+_\Phi(E_q(p))\right),
\end{align}
with the modified distribution functions
\begin{align}
f^+_\Phi(E_p) & =\frac{ \left( \bar\Phi + 2\Phi \expm \right) \expm +
  \expmmm } {1 + 3\left( \bar\Phi + \Phi \expm \right) \expm +
  \expmmm}, \nom\\
f^-_\Phi(E_p) & =\frac{ \left( \Phi + 2 \bar\Phi \expp \right) \expp +
  \expppp }{1 + 3\left( \Phi + \bar\Phi \expp \right) \expp +
  \expppp}. \nom
\end{align}

\section{Results}
\label{sec-res}

\begin{figure}[ht]
  \centering
  \begin{minipage}{.48\textwidth}
  \centering
  \includegraphics[width=1.1\textwidth]{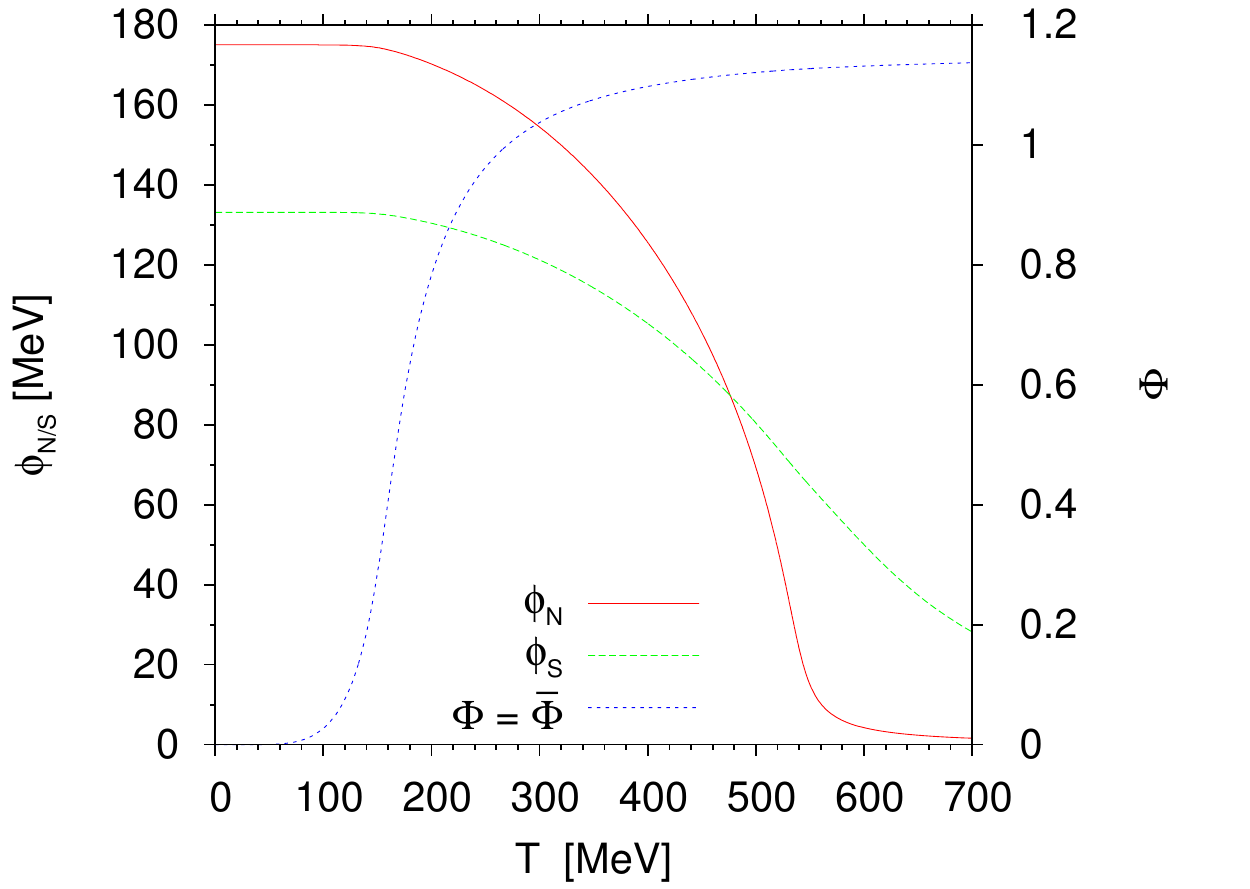}
  \caption{Temperature dependence of the order parameters with
    $m_{\sigma}=1.3$ GeV}
  \label{fig-fields_high_sigma}
\end{minipage}
\hspace*{0.02\textwidth}
\begin{minipage}{.48\textwidth}
  \centering
  \includegraphics[width=1.1\textwidth]{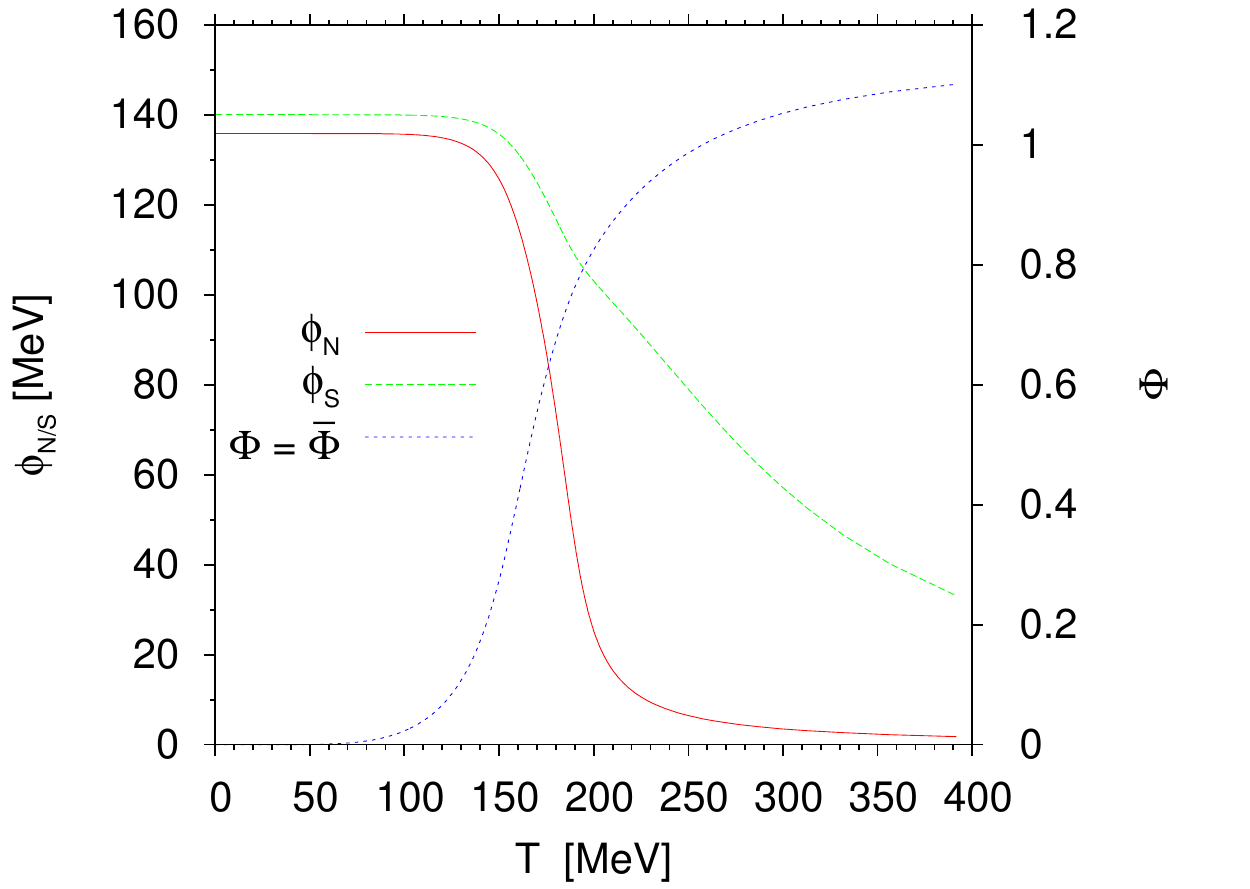}
  \caption{Temperature dependence of the order parameters with
    $m_{\sigma}=0.4$ GeV}
  \label{fig-fields_low_sigma}
\end{minipage}
\end{figure}

Solving Eqs.~\ref{eq_Phi}-\ref{eq_phiS} we get the temperature
dependence of the order parameters, which can be seen in
Fig.~\ref{fig-fields_high_sigma}. 
In \cite{Parganlija_2013} it was shown that the $q\bar{q}$ scalar
nonet most probably contains $f_0$'s with masses higher than
$1$~GeV. If we set $\lambda_1=0$ we get $m_{f_0^L} = 1.3$~GeV, which
is in agreement with \cite{Parganlija_2013}. However in this case we
get a very high pseudocritical temperature, $T_c\approx 550$~MeV, for
$\phi_N$, which is much larger than earlier results (e.g. on lattice
$T_c\approx 150$~MeV \cite{Aoki}). Now, if we tune $\lambda_1$ to get
$m_{f_0^L} 400$~MeV (which corresponds to the physical particle
$f_0(500)$), than $T_c$ goes down to $150-200$~MeV, which can be
seen in Fig.~\ref{fig-fields_low_sigma}. This suggests that in order
to get a good pseudocritical temperature we would need a
scalar-isoscalar particle with low mass ($\sim 400$~MeV),
which is most probably not a $q\bar{q}$ state according to
\cite{Parganlija_2013}.

\begin{acknowledgement}
Authors were supported by the Hungarian OTKA fund K109462.
\end{acknowledgement}


\end{document}